\documentclass[letter]{aa} 
\usepackage{graphicx}
\usepackage{txfonts}
\def\Lsun{L_{\sun}}

\def\Lstar{L_{\star}}

\def\Tstar{T_{\star}}

\def\epsC{\varepsilon_{\rm C}}

\def\epsCfree{\tilde{\varepsilon}_{\rm C}}
\def\fcfreeC{\tilde{f}_{\rm C}}
\def\ffreeC{F_{\rm C}}

\def\cm2g{{\rm cm^2 / g}}

\begin{document}
   \title{Winds of M- and S-type AGB stars: an unorthodox suggestion for the driving
   mechanism}

   \titlerunning{On the wind mechanism of M- and S-type AGB stars}


   \author{S.\,H{\"o}fner\inst{1}
          \and
          A.C.\,Andersen\inst{2}
          }

   \offprints{S.\,H{\"o}fner}

   \institute{Department of Astronomy and Space Physics, Uppsala University,
              Box 515, SE-751\,20 Uppsala, Sweden\\
              \email{hoefner@astro.uu.se}
         \and
             Dark Cosmology Centre, Niels Bohr Institute, University of Copenhagen,
         Juliane Maries Vej 30, DK-2100 Copenhagen, Denmark\\
             \email{anja@dark-cosmology.dk}
             }

   \date{Received <date>; accepted <date> }


  \abstract
   {Current knowledge suggests that the dust-driven wind scenario
   provides a realistic framework for understanding mass loss from C-rich
   AGB stars. For M-type objects, however, recent detailed models
   demonstrate that radiation pressure on silicate grains is not sufficient
   to drive the observed winds, contrary to previous expectations.}
   {In this paper, we suggest an alternative mechanism for the mass-loss of
   M-type AGB stars, involving the formation of both carbon and silicate grains due to
   non-equilibrium effects, and we study the viability of this scenario.}
   {We model the dynamical atmospheres and winds of AGB stars
   by solving the coupled system of frequency-dependent radiation hydrodynamics and
   time-dependent dust formation, using a parameterized description
   of non-equilibrium effects in the gas phase.
   This approach allows us to assess under which circumstances it is possible to
   drive winds with small amounts of carbon dust and to get silicate grains
   forming in these outflows at the same time.}
   {The properties of the resulting wind models, such as mass loss rates and
   outflow velocities, are well within the observed limits for M-type AGB stars.
   Furthermore, according to our results, it is quite unlikely that significant amounts
   of silicate grains will condense in a wind driven by a force totally unrelated
   to dust formation, as the conditions in the upper atmosphere and wind acceleration
   region put strong constraints on grain growth.}
   {The proposed scenario provides a natural explanation for the observed similarities
   in wind properties of M-type and C-type AGB stars and implies a smooth transition
   for stars with increasing carbon abundance, from solar-composition to C-rich AGB stars,
   possibly solving the long-standing problem of the driving mechanism for stars with
   C/O close to one.}


   \maketitle
%

\section{Introduction}\label{s_intro}

Low- and intermediate-mass stars are known to lose a significant
fraction of their mass through slow, massive winds during the
Asymptotic Giant Branch (AGB) phase of their evolution. The basic
scenario for this process is that of pulsation-enhanced dust-driven
winds: shock waves created by stellar pulsation lead to a dense,
cool extended stellar atmosphere, allowing for efficient dust
formation. The grains are accelerated away from the star by
radiation pressure, dragging gas along. The composition of the
grains is determined by the elemental abundances in the stellar
atmosphere, with C-rich stars forming carbon grains, and O-rich
environments producing silicate particles.

This scenario seems to work well for the case of C-rich AGB stars,
as demonstrated by comparison of detailed self-consistent dynamical
models with observations as diverse as low-resolution IR spectra
(e.g., H{\"o}fner et al.\ 2003, Gautschy-Loidl et al.\ 2004) and
line profile variations of CO vibration-rotation lines (e.g.,
Nowotny et al.\ 2005).

Recent work on similar models for stars with C/O$\,<1$, however,
faces serious problems (Woitke 2006; H{\"o}fner 2007).
Frequency-dependent dynamical models demonstrate that the opacities
of silicate grains are too low to drive a wind, even when assuming
rather favorable grain compositions. In addition, these models show
that the Fe-content of the grains has to be extremely low, as the
radiative equilibrium temperature of silicate grains increases
strongly with increasing inclusion of Fe, which leads to even lower
grain opacities than previously assumed. Therefore, the very
ingredient that makes C-rich models more successful than ever, i.e.
frequency-dependent radiative transfer, prevents dust driven winds
in stars with C/O $< 1$.

While alternative wind mechanisms have been discussed earlier,
observations indicate that winds of O-rich AGB stars behave very
similar to their C-rich counterparts. Furthermore, the very fact
that silicate grains are formed in such outflows puts rather strict
constraints on the possible mechanisms, as will be discussed in
Sec.~\ref{s_disc}. This leads us to propose a scenario, where the
winds of M- (and probably S-) type AGB stars are actually driven by
small amounts of carbon grains, with silicates forming as a
by-product. This requires that a certain fraction of the carbon
atoms in the dust formation region are available for grain
formation, i.e. not bound up in CO.

This scenario implies deviations from chemical equilibrium (CE) in
the gas phase, probably brought about by strong atmospheric shock
waves, which seems not too far-fetched in the light of recent
observations showing the existence of molecules typically expected
in O-rich environments in C-rich stars (e.g., Sch{\"o}ier et al.\
2006), and vice versa. Current models of non-equilibrium gas-phase
chemistry by Cherchneff (2006) predict rather small deviations from
CE (less than what we assume below), but these models are computed
for significantly higher densities (which should favor more CE-like
compositions), and without taking condensation into account. As an
inclusion of detailed non-CE in the gas phase in our wind models is
well beyond the scope of this first investigation, we use the
abundance of available carbon as a parameter of the models. This
approach allows us to test under which circumstances it is possible
to drive winds {\it and} to get a significant amount of silicate
grains forming in the outflows.


\section{The hydrodynamical model}\label{s_dynmod}

Our hydrodynamical models predict mass-loss rates and wind
velocities of AGB stars, as well as the amount of dust formed in the
winds, by treating in detail the atmosphere and the circumstellar
environment around pulsating long-period variable stars. This is
done by solving the coupled system of frequency-dependent radiation
hydrodynamics and time-dependent dust formation (cf. H{\"o}fner et
al.\ 2003). The calculations presented here are based on opacity
sampling data of molecular opacities
at 64 frequency points between 0.25 and 25 $\mu$m.

\begin{table*}[ht]
\caption{Model parameters and input data: luminosity ${L_{\star}}$,
effective temperature ${T_{\star}}$ (initial model; mass
${M_{\star}} = 1 {M_{\odot}}$ for all models); fraction of free
carbon $F_{\rm C}$; period $P$ and velocity amplitude of piston
${\Delta u_{\rm p}}$ (luminosity factor $f_L = 2.0$ for all models,
see Gautschy-Loidl et al. 2004); opacity data for amorphous carbon
$\kappa_{\rm C}$: RM = data from Rouleau \& Martin (1991, sample AC2), 
J4 = data from J{\"a}ger et al. (1998, sample cel400). Resulting wind
properties: mass loss rate $\dot{M}$, mean velocity at the outer
boundary ${\langle u \rangle}$, degree of condensation of silicon
$f_{\rm Si}$, of free carbon ${\tilde{f}_{\rm C}}$ and of all carbon
$f_{\rm C}^{\rm tot}$.} \label{t_mod}
\begin{center}
\begin{tabular}{lllcllcclll}
\hline
\hline
%
 ${L_{\star}}$ $[{L_{\odot}}]$ & ${T_{\star}} [K]$  & $F_{\rm C}$ & ${\kappa_{\rm C}}$ & $P$ [d] & ${\Delta u_{\rm p}} [km/s]$
  & ${\dot{M}}$ [${M_{\odot}}$/yr] & ${\langle u \rangle}$  [km/s] & ${f_{\rm Si}}$ & ${\tilde{f}}_{\rm C}$ & ${f^{\rm tot}_{\rm C}}$
 \\
\hline
\hline
%
5000 & 2800 & 0.7 & RM & 310 & 6.0 & $ 5 \cdot 10^{-7}$ &  4 & $ 3 \cdot 10^{-2}$ & 0.3 & 0.2 \\
7000 & 2700 & 0.4 & RM & 390 & 6.0 & $ 2 \cdot 10^{-6}$ &  4 & $ 5 \cdot 10^{-2}$ & 0.4 & 0.2 \\
7000 & 2700 & 0.5 & RM & 390 & 6.0 & $ 3 \cdot 10^{-6}$ &  7 & $ 1 \cdot 10^{-2}$ & 0.4 & 0.2 \\
10000 & 2600 & 0.3 & RM & 525 & 6.0 & $ 1 \cdot 10^{-5}$ &  6 & $ 4 \cdot 10^{-2}$ & 0.5 & 0.2 \\
10000 & 2600 & 0.3 & RM & 525 & 3.0 & $ 5 \cdot 10^{-6}$ &  6 & $ 2 \cdot 10^{-1}$ & 0.5 & 0.2 \\
10000 & 2600 & 0.4 & RM & 525 & 3.0 & $ 7 \cdot 10^{-6}$ & 10 & $ 2 \cdot 10^{-2}$ & 0.4 & 0.2 \\
%
\hline
7000 & 2700 & 0.4 & J4 & 390 & 6.0 & $ 1 \cdot 10^{-6}$ &  4 & $ 3 \cdot 10^{-1}$ & 0.3 & 0.1 \\
10000 & 2600 & 0.3 & J4 & 525 & 3.0 & $ 4 \cdot 10^{-6}$ &  6 & $ 5 \cdot 10^{-1}$ & 0.4 & 0.1 \\
10000 & 2600 & 0.4 & J4 & 525 & 3.0 & $ 6 \cdot 10^{-6}$ & 10 & $ 6 \cdot 10^{-2}$ & 0.4 & 0.2 \\
\hline
\hline
\end{tabular}
\end{center}
\end{table*}

In contrast to our earlier models, we have implemented a
time-dependent description of grain growth for silicate grains
(H{\"o}fner et al., in prep.), and we focus on the formation of pure
forsterite particles (Mg$_2$SiO$_4$), as non-grey effects will force
a low Fe-content (see Sect.\,~\ref{s_intro}). While treating the
growth of grains in full non-equilibrium, we do not consider the
nucleation of new grains from the gas phase but assume the presence
of seed nuclei (a given number per H atom) at the point where grain
growth becomes possible. Opacity data for the relevant grain
materials is taken from J{\"a}ger et al.\ (2003; forsterite),
Rouleau \& Martin (1991; amorphous carbon sample AC2) and J{\"a}ger
et al.\ (1998; amorphous carbon sample cel400).

All models discussed here have solar elemental abundances (i.e. C/O
$\approx 0.5$), but we assume that a certain fraction of carbon is
not bound in CO in the region relevant for grain growth. This
fraction is parameterized by a factor $\ffreeC$ = $\epsCfree /
\epsC$, where $\epsCfree$ is the abundance of carbon available for
dust formation. This is admittedly a very simple-minded treatment of
non-equilibrium in the gas phase, but it allows us to assess, how
much free carbon is needed to drive a wind for given stellar
parameters. Note also, that only a fraction of this fraction of all
carbon atoms will actually condense into grains (see
Tab.\,~\ref{t_mod} and discussion below).

\section{Results and discussion}\label{s_disc}

We have calculated a number of models with different combinations of
stellar parameters, fractions of free carbon ($\ffreeC$), and
pulsation amplitudes, using two different sets of opacity data for
amorphous carbon\footnote{This term actually covers a variety of
materials with different microscopic structures and different
optical properties (see discussion in Andersen et al.\ 2003).}. The
model parameters and the resulting wind properties are summarized in
Tab.\,~\ref{t_mod}. As expected, the fraction of free carbon
required to obtain a dust-driven wind decreases with increasing
luminosity and decreasing effective temperature. Also, in accordance
with previous results on C-rich wind models, the mean degree of
condensation of free carbon $\fcfreeC$ is never close to one in
these models, and the degree of condensation of all carbon (obtained
by multiplying $\fcfreeC$ with $\ffreeC$) is rather low in certain
cases.

\begin{figure*}
\centering
\vspace{0.5 cm}
\includegraphics [width=14cm, angle=270] {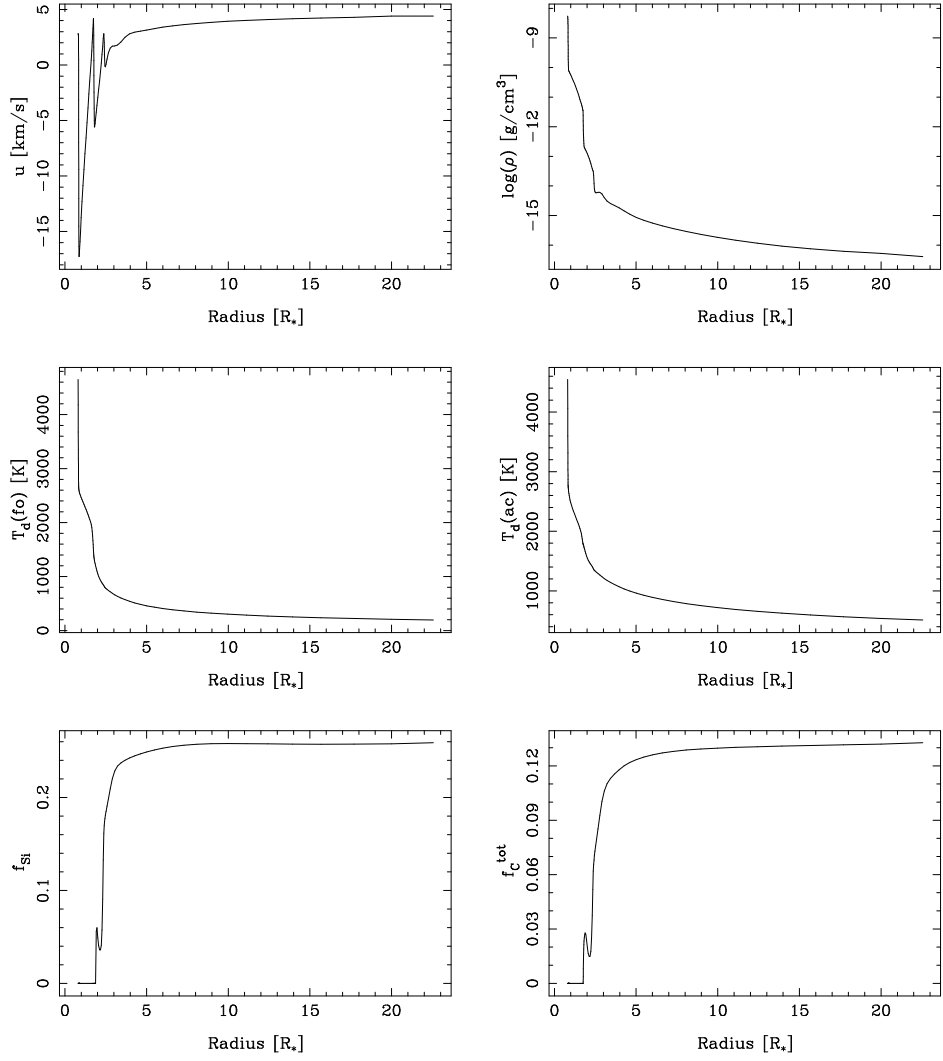}
\caption{Snapshot of the radial structure of the model with $\Lstar
= 7000 \, \Lsun$, $\Tstar = 2700 \,$K and opacity data for amorphous
carbon by J{\"a}ger et al.\ (1998; sample cel400). Top row: gas
velocity and density; middle row: radiative equilibrium temperature
for forsterite and carbon grains; bottom row: degree of condensation
for Si and C. Radius in units of the stellar radius.} \label{f_rstruct}
\end{figure*}

What may seem more surprising at a first glance, is the fact that
the degree of condensation for Si (and, correspondingly, the amount
of silicate grains formed in the wind) is quite low in most models,
and actually decreases with conditions that make the formation of
carbon grains more efficient (e.g., higher pulsation amplitude, or a
larger fraction of free carbon, for given stellar parameters). The
low silicate dust abundance is not a matter of temperatures being
too high to allow for condensation,
but rather a matter of grain growth timescales: as soon as enough
carbon dust is formed to accelerate the gas away from the star, the
condensation of dust (and in particular of silicate grains,
handicapped by lower abundances of critical elements like Si and Mg)
becomes a race against falling densities in the outflow, a
phenomenon that finally stops condensation, both for silicates and
carbon grains (leading to $\fcfreeC < 1$), despite falling gas and
grain temperatures.

Therefore it is actually quite remarkable that silicate
grains can form at all in a wind driven by some other force (e.g.
radiation pressure on another, unrelated dust species as in this
case), as the rapid slowing down of grain growth in an outflow puts
tight constraints on the wind mechanism. If the silicate grains do
not form simultaneously with (or maybe even prior to) the driving
dust species (carbon), i.e. at the same distance from (or maybe even
closer to) the star, at rather high gas densities, there is
virtually no chance of their forming at all.

In order to understand how grains consisting of such different
materials as amorphous carbon and silicates may form at basically
the same distance from the star, we have to take a look at two types
of temperatures: the condensation temperature of a material (setting
the threshold for grain formation) and the grain temperature at a
given distance from the star, which, in practice, will be equal to
the radiative equilibrium temperature corresponding to the grain
material. At the densities prevailing in the pulsating atmosphere,
amorphous carbon grains will typically start to form around
1500-1800\,K while the formation of forsterite requires temperatures
as low as 1000-1200\,K. However, due to their optical properties
(opacity decreasing with wavelength around stellar flux maximum),
carbon grains will be hotter than the black body temperature at a
given distance (cf.\ H{\"o}fner et al.\ 2003, Fig.~6), while
forsterite grains will be cooler (opacity increasing with wavelength
in the critical regime).\footnote{At the relevant wavelengths, the
opacity of amorphous carbon grains is roughly proportional to
$1/\lambda$. Therefore the carbon grains absorb and emit radiation
preferentially at short wavelengths. This property in combination
with the radiative equilibrium condition requires the grains to be
hotter than the radiation temperature in order to emit as much
energy as they absorb.}  This means that, at the same distance from
the star, forsterite grains will be significantly cooler than
amorphous carbon grains, and the critical condensation temperature
is reached approximately in the same region for both types of dust
(Fig.~\ref{f_rstruct}).

The strong sensitivity of a non-driving dust species (i.e. silicate
grains in the present case) to the conditions in the wind
acceleration region is illustrated by a comparison of models
calculated with two different opacities for the carbon grains, but
otherwise identical parameters. The data of J{\"a}ger et al.\ (1998;
{\it cel400}) has a steeper slope in the region around 1$\, \mu$m,
than many other types of amorphous carbon (cf.\ Fig.\,2 in Andersen
et al.\ 2003) leading to a higher grain temperature at a given
distance. Therefore, the formation of carbon grains (and,
consequently the wind acceleration region) is shifted slightly
outwards in these models compared to the ones based on the Rouleau
\& Martin (1991) data, improving the conditions for the formation of
silicate grains. This results in an increased degree of condensation
of Si, while the mass loss rate and wind velocity (governed by the
carbon grains) are basically unchanged (cf.\
Tab.\,~\ref{t_mod})\footnote{Note, again, that the degree of
condensation of Si is a direct measure of the amount of silicate
grains formed in the wind.}. These results indicate that it may be
very difficult to form silicate grains in a wind driven by a force
totally unrelated to dust formation, as the conditions in the upper
atmosphere and wind acceleration region put strong constraints on
grain growth.

In addition to these theoretical considerations, several existing
observations can be interpreted in favor of the proposed scenario:
Sch{\"o}ier et al. (2006) have recently demonstrated the existence
of molecules typical of O-rich environments in C-rich AGB stars, a
finding that is hard to reconcile with standard chemical equilibrium
models. Interferometric and spectroscopic observations indicate the
existence of very dense layers of water vapor in O-rich stars (e.g.,
Weiner 2004, Ohnaka 2005, Justtanont 2005), potentially requiring
abundances of H$_2$O in excess of what can be expected from chemical
equilibrium (where an amount of oxygen equal to the total amount of
carbon is bound in CO). Furthermore, while the wind properties of
C-rich and O-rich AGB stars are very similar, there is an average
tendency of slightly higher outflow velocities at given mass loss
rate for C-rich stars (cf., e.g., Olofsson 2004). This could be a
natural consequence of all these winds being driven by carbon
grains, but with an increasing amount of carbon being available for
condensation, going from O-rich to C-rich objects.

\section{Conclusions}

We have investigated the viability of a scenario where winds of
O-rich AGB stars are driven by small amounts of carbon grains. Our
approach allows us to estimate both lower and upper limits for the
required non-equilibrium effects in the gas, with one end of the
spectrum being defined by the minimum amount of carbon dust
necessary to drive an outflow, and the other by the decreasing
amounts of silicates being formed in the wind with increasing
availability of carbon. Our model shows that significant amounts of
forsterite grains (recently found in comets by STARDUST) can form in
such winds.

The proposed scenario provides a natural explanation for the
observed similarities in wind properties of O-rich and C-rich cool
giants and implies a smooth transition for stars with increasing
carbon abundance, from M-type to C-type AGB stars, probably solving
the long-standing problem of the driving mechanism for stars with
C/O close to one (S-type AGB stars; cf., e.g., Ramstedt et al.
2006).

Crucial, if non-trivial, tests of the scenario can be provided by
observations which can assess the relative abundances of molecules
in the critical dust formation region, in order to check for
non-equilibrium effects in the gas phase. A direct detection of
carbon grains in the spectra is rather unlikely: the optical
properties of amorphous carbon lead to remarkably featureless IR
spectra. SiC, on the other hand, which is often used as an indicator
for dust in C-rich stars due to its pronounced feature around $11
\mu$m will hardly be produced in sufficient quantities to be
detected above the background of amorphous carbon and the nearby
silicate feature. Considering the relative opacities of SiC and
silicates, this would require at least as much Si being bound in SiC
as in silicates, which seems unlikely in an environment with C/O $<
1$. 

\begin{acknowledgements}
SH acknowledges support form the Swedish Research
Council ({\it Vetenskapsr{\aa}det}\,).
The Dark Cosmology Centre is funded by the Danish National Research
Foundation.
\end{acknowledgements}

 \end{document}